# Lightweight Ciphers Based on Chaotic Map – LFSR Architectures

M. Garcia-Bosque, C. Sánchez-Azqueta, S. Celma
Group of Electronic Design
Universidad de Zaragoza
Zaragoza, Spain
{mgbosque, csanaz, scelma}@unizar.es

*Abstract*—In this paper, we propose and analyze two different stream ciphers based on a Skew Tent Map and a Modified Logistic Map respectively. In order to improve the randomness of these systems, a single method for increasing the period length of the generated sequences has been applied. The results prove that the randomness of these systems can be severally increased by using this method, making these systems suitable for secure communications.

## I. Introduction

In the last years, it has become necessary to encrypt high amounts of data (e.g. video) in real time for different applications. Usually, block ciphers (such as DES, AES, RC5, etc.) or stream ciphers (such as RC4, A5/1, A5/2, HC-356, etc.) are used to private data encryption due to their high speed and low complexity. However, these algorithms are not able to encrypt high amounts of data in real time or are not secure enough [1]. Although some methods have been proposed to increase the encryption speed of these systems, the security is negatively affected. Thus, there is a big interest in finding new algorithms that are both fast and secure.

In this context, chaos-based cryptosystems have emerged as a promising alternative to classical encryption since they could be able to achieve a good balance between speed and security [2]. Most of these cryptosystems are based on one dimensional chaotic maps since they are the simplest systems that present a chaotic behavior. Specially, the logistic map has been used in many of the proposed cryptosystems due to its simplicity, high throughput and ergodic properties.

Unfortunately, most of these systems have problems at the moment of defining the space key. Ideally, in a cryptosystem, all keys should be equally strong. However, in most of the "secure communication systems" proposed, there are some keys that lead to non-uniform sequences, due to the presence of periodic windows. A bifurcation diagram can help a designer to find the intervals in which a given interval generates periodic windows, in order to avoid them. However, in many chaotic systems, even within the chaotic region there exist periodic windows unsuitable for secure communications. As an example, the logistic map, used in many secure communication systems, have periodic windows in the chaotic region. Although this system has been widely studied, the parameters that generate these periodic windows are not fully known.

In order to have a chaotic system suitable for secure communications, it is preferred to have a continuous region where all the parameters give rise to complete chaoticity. For this reason, we have chosen the Modified Logistic Map and the Skew Tent Map for our cryptosystems.

In this paper, we propose and analyze a two different secure communication systems based on a Modified Logistic Map and a Skew Tent Map respectively. These maps are combined with a Linear Feedback Shift Register in order to increase the randomness of the generated sequences.

The paper is organized as follows: Section II presents the chaotic maps that have been used; Section III explains the proposed encryption algorithms in detail; Section IV presents the simulation results and a cryptanalysis of our communication systems; finally, the conclusions are drawn in Section V.

## II. Skew Tent and Modified Logistic Maps

### A. Skew Tent Map

The Skew Tent Map (STM) is expressed mathematically as:

$$f(x_i) = x_{i+1} = \begin{cases} x/\gamma, & x \in [0, \gamma] \\ (1-x)/(1-\gamma), & x \in (\gamma, 1] \end{cases} \quad (1)$$

where $\gamma, x_0 \in (0,1)$. For any value of $\gamma$ and $x_0$ the map will be chaotic [3] and, therefore, will be suitable to be used in chaos-based cryptosystems.

### B. Modified Logistic Map

The classic logistic map is expressed mathematically as:

$$f(x_i) = x_{i+1} = \gamma x_i (1 - x_i), \quad x \in [0,1] \quad (2)$$

This map presents chaos for most values of $3.57 < \gamma \leq 4$ while, for $\gamma > 4$, most initial values eventually leave the interval $[0,1]$ and diverge.

This map, however, has some issues that prevents it from being implemented directly in cryptosystems. First, as it has been said before, there is an open and dense set of parameters $\gamma$ for which the map is regular (i.e. there exist periodic windows). It is necessary to avoid using these parameters $\gamma$ in order to generate truly chaotic sequences.

Furthermore, the images are not uniformly distributed in the interval $[0,1]$. Instead, it can be easily proved that the images are

This work has been partially supported by MINECO-FEDER (TEC2011-23211 and TEC2014-52840-R) and FPU fellowship (FPU14/03523).

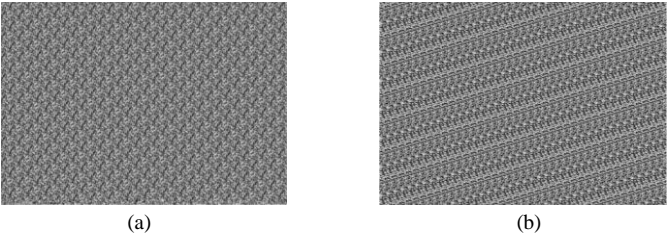

(a)  (b)

Fig. 1. (a) Binary sequence generated by the STM and (b) Binary sequence generated by the MLM.

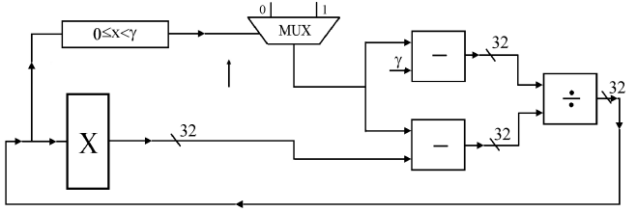

Fig. 2. Block diagram of a 32 bits Skew Tent Map generator.

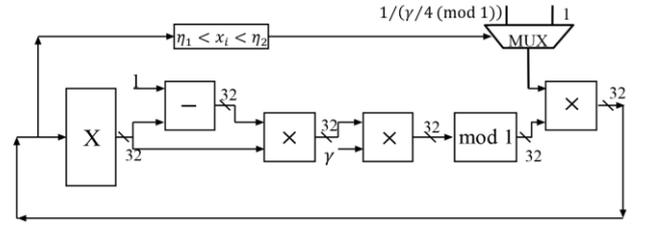

Fig. 3. Block diagram of a 32 bits Modified Logistic Map generator.

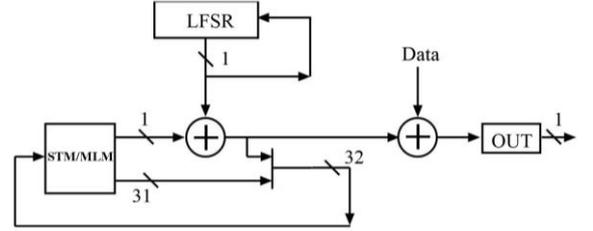

Fig. 4. Block diagram of the proposed cipher.

distributed in the interval $[0, \gamma/4]$. Therefore, by measuring the length of the attractor, the value of $\gamma$ can be determined. This can be a problem if the value of $\gamma$ is being used as part of the key. In this paper, we will use a Modified Logistic Map (MLM) proposed in [4] that is capable of solving these issues. The MLM is defined by

$$f(x_i) = x_{i+1} = \begin{cases} \gamma x_i(1-x_i) \pmod 1, & x \in I_{ext} \\ \dfrac{\gamma x_i(1-x_i) \pmod 1}{\frac{\gamma}{4} \pmod 1}, & x \in I_{int} \end{cases} \quad (3)$$

with $I_{ext} \in (0,1)\setminus I_{int}$, $I_{int} = [\eta_1, \eta_2]$, $\eta_1 = (1/2) - \sqrt{(1/4) - [\gamma/4]/\gamma}$ and $\eta_2 = (1/2) + \sqrt{(1/4) - [\gamma/4]/\gamma}$ where $[\gamma/4]$ is the greatest integer minor or equal to $\gamma/4$.

It has been proved in [4] that, as long as $\gamma \gtrsim 3,57$, this map has a chaotic behavior and is able to generate numbers uniformly distributed between 0 and 1.

III. ENCRYPTION ALGORITHMS

A. *Increasing the period length*

When a chaotic map is digitalized using $n$ bits, the sequence generated becomes periodic and its maximum period length is $2^n$. Therefore, a digitalized map will never be strictly chaotic. Furthermore, the period lengths of the sequences generated by the STM and the MLM are usually much shorter than this value, which results in poor randomness. As an example, Fig. 1 (a) represents in a matrix of black and white pixels the bits obtained from a particular short period sequence generated by the Skew Tent Map using a 32-bits implementation. An analogue representation for a particular sequence generated by the Modified Logistic Map is shown in Fig. 1 (b). As it can be seen, both these sequences are far from random. Furthermore, another critical example occurs when, due to rounding errors, the number 0 is generated and, therefore, the following numbers of the sequence generated using (1) or (3) are also 0.

Our proposed communication systems are capable of solving these issues by adding a small perturbation $\Delta x_i$ to each element of the sequence (i.e. $\tilde{x}_i = x_i + \Delta x_i$), in order to emulate the trajectory instability of chaotic systems. Using this method, the period length of the generated sequences can be set up to a minimum value, using the following proposition proven in [5].

*Proposition 1:* Let us consider the binary sequences $y(p)$, $z(p)$ and $w(p) = y(p) \oplus z(p)$, for $p \in \mathbb{N}$ and $P_y, P_z, P_w$ their respective periods. If $P_z > 1$ is prime, then $P_w = mP_z$ with $m \geq 1$.

Using this proposition, we can set $w(p)$ as the least significant bit (LSB) of $\tilde{x}_i$, $y(p)$ as the LSB of $x_i = f(x_{i-1})$ and $z(p)$ as the output bits of a pseudo-random sequence with a prime period $P_z$. Then the perturbed trajectories of $\tilde{x}$ will have a period $P_w \geq P_z$.

In our work, we propose to use a Linear Feedback Shift Register (LFSR) in order to perturb the sequences generated by the MLM or the STM. In order to guarantee that the period of the generated sequences is big enough, we must use an appropriate order $k$ so that the period of the LFSR, $P_z = 2^k - 1$, is a prime number (Mersenne prime). Some possible values of $k$ are 19, 31, 61 or 89.

B. *Proposed communications systems*

The encryption algorithms generate pseudo-random sequences $\tilde{x}$ by perturbing the orbits generated by our chaotic system (STM or MLM) with an LFSR. To that end, the LSB of each $x_i$ is XORed with a bit generated by a LFSR. The resulting sequence of bits is combined with the plaintext using an XOR gate. On the other hand, the perturbed $\tilde{x}_i$ is used to generate the next number $x_{i+1}$ using (1) or (3). The block diagram of each chaotic generator is shown in Fig. 2 and Fig. 3. The whole encryption system used in the transmitter is shown in Fig. 4.

The receiver is also made of a chaotic map (STM or MLM) and an LFSR. By using identical initial values $(x_0, \gamma)$, it generates the same pseudorandom sequence that was produced in the transmitter. By combining this sequence with the encrypted message with another XOR gate, the original message is recovered.

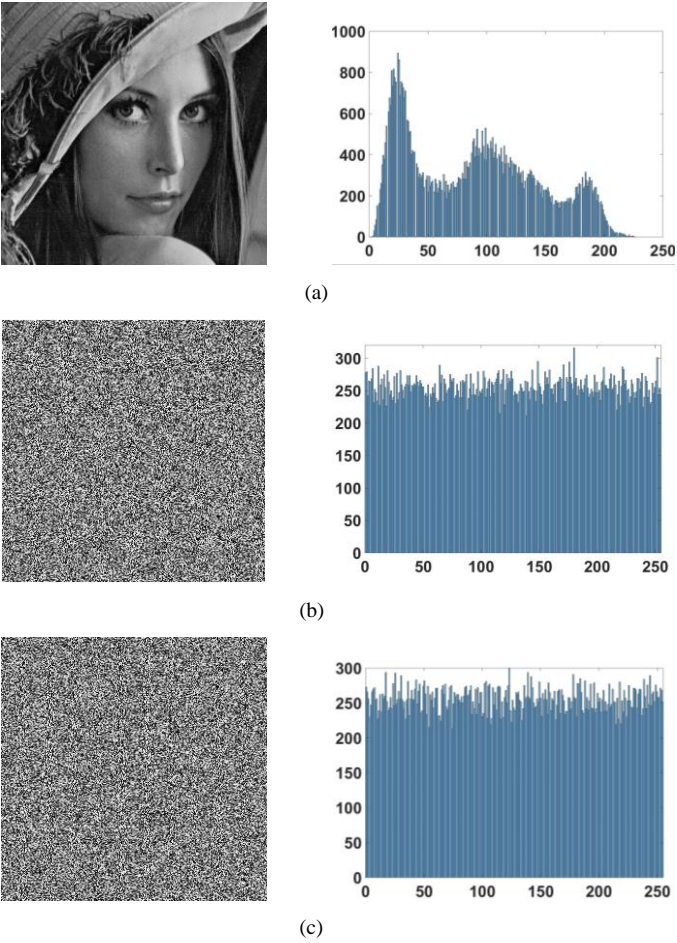

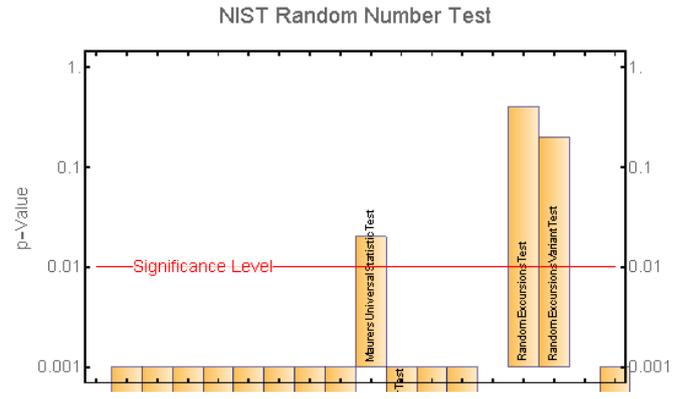

Fig. 6. NIST tests results for a sequence generated using only the STM.

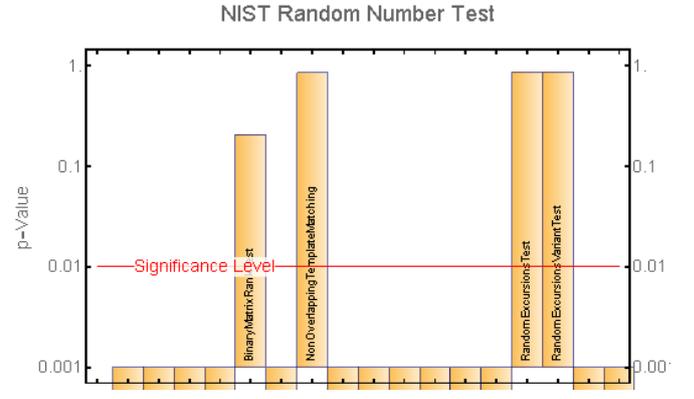

Fig. 7. NIST tests results for a sequence generated using only the MLM.

Fig. 5. (a) Test image, (b) encrypted image using the STM-LFSR algorithm and (c) encrypted image using the MLM-LFSR algorithm with their respective histograms.

## IV. SIMULATION RESULTS AND CRYPTANALYSIS

The operation of these communication systems has been simulated with Matlab software. For the simulation tests, a single precision (32 bits) arithmetic has been used. A 31-order LFSR has been used in each system to guarantee that the period of each cycle is at least $P_w \geq 2^{31} - 1 \approx 2.1 \times 10^9$.

Both of these systems have been able to encrypt the images efficiently as shown in Fig. 5. The correlation between the original and encrypted images have been calculated using the "corr2" function of Matlab. The result are -0.0041 for the STM-LFSR system and -0.0027 for the MLM-LFSR system. As it can be seen, the correlation in both systems is very low, indicating that both systems are able to encrypt the images. However, for the MLM based system, the correlation is a bit smaller, which suggests that this system might be better for image encryption.

### A. Randomness test

In order to analyze the security of these algorithms, several sequences have been encrypted and subjected to the National Institute of Standards and Technology (NIST) SP 800-22 battery of test [6]. Furthermore, we have tested some sequences generated using only the STM and the MLM to verify if our method used to increase the period length does in fact increase the randomness of the generated sequences.

After analyzing the sequences we have observed that none of the sequences generated using only the STM or the MLM have passed all of the NIST tests (Fig. 6 and Fig. 7). However, all of the sequences generated by the proposed algorithms have been able to pass the NIST test. As an example, Fig. 8 and Fig. 9 show the NIST results for a particular sequence generated by each algorithm. It can clearly be seen that both sequences have passed the NIST tests.

### B. Key space

One of the aspects that need to be studied in any cryptosystem is the key space.

First, the key space size should be big enough to prevent an attacker to brute-force the key. In our algorithms, the key is composed by the initial value $x_0$, the control parameter $\gamma$ and the initial state of the LFSR. Therefore, assuming that $x_0$ and $\gamma$ have a 32-bit precision, the total number of possible keys would be $\kappa = 2^{32+32+31} = 2^{95}$. However, depending on the implementation, this number is usually a bit smaller since $x_0$ and $\gamma$ are constrained in a certain region. For example, in the STM, $x_0$ and $\gamma$ must be between 0 and 1 and, in the IEEE 754 single precision format, there are only $\sim 2^{30}$ possible values in that range which would result in $2^{91}$ possible keys.

Traditionally, a key-space of $\kappa > 2^{80}$ has been considered to be secure and it is still considered big enough for the coming years [7]. Therefore, our key space is big for current applications

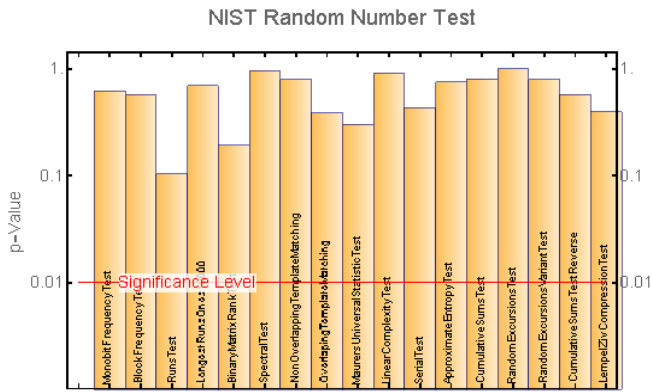

Fig. 8. NIST results for a sequence generated using the STM-LFSR system.

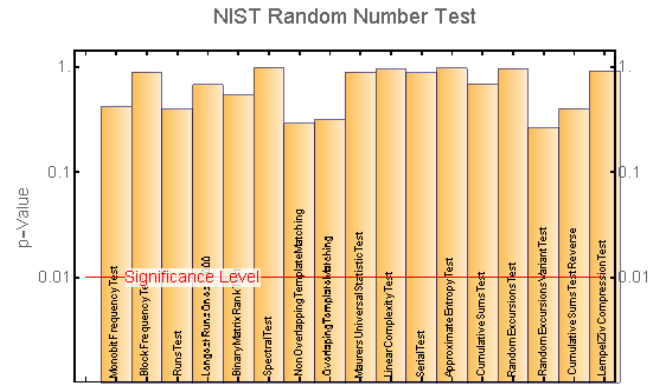

Fig. 9. NIST results for a sequence generated using the MLM-LFSR system

However, for long term applications, the key size should be increased to, at least, $\kappa > 2^{112}$ in order to prevent the system to be brute-forced [8]. This could be easily achieved by implementing our algorithms using double precision (64 bits) although the encryption speed would decrease.

In the second place, all of the possible keys should be equally secure since, otherwise, an attacker could decrypt a message encrypted with a weak key. As it has been explained above, both of our algorithms work in a chaotic regime and use an LFSR to avoid any possible periodic windows. Therefore, it can be concluded that generated sequences will be equally secure regardless of the key used.

Finally, the generated sequences must be very sensitive on the key to prevent the attackers to find relationship between the keys and their corresponding masking sequences. This property is also satisfied since, the usage of a chaotic system causes that similar keys generate very different sequences.

*C. Sensitivity dependence on the plaintext*

Typically, another important property that is required in a cryptosystem is that a single change in the plaintext generates a completely different ciphertext. This property is not satisfied in our cryptosystems since, like in many other stream ciphers, a single bit change in the plaintext only causes a single bit change in the ciphertext. This makes these systems vulnerable to some attacks such as differential known-plaintext attack if several messages are encrypted using the same key. Therefore, in order to be secure it should be avoided to encrypt several messages with the same key.

V. CONCLUSIONS

Two different stream ciphers have been proposed and analyzed. Both of them have passed the NIST tests, proving the good randomness of the generated sequences. Furthermore, a cryptanalysis has been presented to prove that these systems are secure. However, although both cryptosystems are similar, there are some differences that are worth noticing in order to choose between them.

First, when we have analyzed the correlation between the original and the encrypted images, we have found that the correlation is a bit smaller when the MLM-LFSR system has been used. Furthermore, although both of the communication systems have passed the NIST randomness test, the MLM-LFSR communication system seems to offer slightly better results during the tests. On the other hand, the MLM can be more difficult to implement, due to the mod 1 calculation.

In conclusion, both systems could be suitable for applications that require both speed and security. However, it is difficult to determine which algorithm is better. In order to choose between them, the application and the platform where it is going to be implemented should be considered.

These ciphers are currently being implemented in a Zedboard development board that includes a Zynq-7020 SoC. The implementation results including the total number of LUTs and the throughput will be presented in the conference.

REFERENCES

[1] J. Shah and V. Saxena, "Video Encryption: A Survey," International Journal of Computer Science Issues, vol. 8, no. 2, pp. 525-534, March 2011.

[2] R. Hasimoto-Beltran, "High-performance multimedia encryption based on chaos," Chaos: Interdisciplinary J. Nonlinear Sci., vol. 18, no. 2, pp. 023110-1 -023110-8, 2008

[3] A. Baranovsky, and D. Daems, "Design of One-Dimensional Chaotic Maps with Prescribed Statistical Properties," International Journal of Bifurcation and Chaos, vol. 5, no. 6, pp. 1585-1598, 1995.

[4] S.L. Chen, S.M. Chang, W.W. Lin, and T.Hwang, "Digital secure-communication using robust hyper-chaotic systems," International Journal of Bifurcation and Chaos, vol. 18, no. 11, pp. 3325-3339, 2008.

[5] L. Kocarev, and S. Lian, "Chaos-based cryptography", Springer, 2009.

[6] NIST Special Publication 800-22 Rev.1a: A statistical test suite for random andpseudorandom number generators for cryptographic applications (April 2010).

[7] ENISA, "Algorithms, key size and parameters report", November 2014. <https://www.enisa.europa.eu/activities/identity-and-trust/library/deliverables/algorithms-key-size-and-parameters-report-2014>

[8] E. Barker, A. Roginsky, "Transitions: Recommendation for transitioning the use of cryptographic algorithms and key lengths". Special publication 800-131 A, NIST, 2011.